\documentclass[12pt]{iopart}
\usepackage{iopams}

\begin{document}
\title{High accuracy results for the energy levels of the molecular ions H$_{2}^{+}$, D$_{2}^{+}$ and HD$^{+}$, up to $J=2$}

\author{J Ph Karr and L Hilico}

\address{D\'epartement de Physique et Mod\'elisation,
Universit\'e d'Evry Val d'Essonne\\
Boulevard F. Mitterrand, 91025 Evry cedex, France}

\address{Laboratoire Kastler Brossel, Universit\'e Pierre et Marie Curie\\
T12, Case 74, 4 place Jussieu, 75252 Paris, France}

\ead{karr@spectro.jussieu.fr}

\submitto{\JPB}

\begin{abstract}
We present a nonrelativistic calculation of the rotation-vibration levels of the molecular ions H$_{2}^{+}$, D$_{2}^{+}$
and HD$^{+}$, relying on the diagonalization of the exact three-body Hamiltonian in a variational basis. The $J=2$ levels
are obtained with a very high accuracy of 10$^{-14}$ a.u. (for most levels) representing an improvement by five orders of
magnitude over previous calculations. The accuracy is also improved for the $J=1$ levels of H$_{2}^{+}$ and $D_{2}^{+}$
with respect to earlier works. Moreover, we have computed the sensitivities of the energy levels with respect to the mass
ratios, allowing these levels to be used for metrological purposes.
\end{abstract}

\section{Introduction}

In recent years, precise calculations in the hydrogen molecular ion H$_{2}^{+}$ and its isotopes have attracted interest,
because these systems appear as promising candidates for the metrology of the electron to proton mass ratio $M_{P}/m_{e}$,
or the ratios of the nuclear masses $M_{D}/M_{P}$ \cite{article1,article2,schiller}. Almost all the rotation-vibration
levels of HD$^{+}$, H$_{2}^{+}$ and D$_{2}^{+}$ including relativistic and radiative corrections have been computed by R.
E. Moss \cite{mosshdplus,mossh2plus,mossd2plus} with an accuracy of 10$^{-9}$ a.u. Recent progress in variational
calculations has allowed to improve the accuracy of the nonrelativistic calculations up to 10$^{-14}$ or even 10$^{-18}$
for the lowest levels \cite{article1,article2,korobovhelium,frolov,yan}, while the accuracy on relativistic and QED
corrections is also improving and should reach $10^{-10}-10^{-11}$ a.u. \cite{korobovhdplus,korobovh2plus}. However, the
optical transitions that may be used in metrology experiments also involve more excited states \cite{article1,article2}
especially in the perspective of comparing several transition frequencies to test the time independence of the mass ratios
as proposed in \cite{schiller}. It would then be useful to extend the high-precision calculations as far as possible into
the rotation-vibration spectrum.

To our knowledge, our method relying on the use of perimetric coordinates is the only one that allows to maintain a high
accuracy up to the most excited vibrational states with reasonable numerical means \cite{article1,article2}. Indeed, this
set of coordinates takes advantage of the dynamical symmetries of the three body Coulomb problem, so that it is possible
to choose a basis in which the Hamiltonian has strict coupling rules, and whose wavefunctions have the correct long-range
behaviour. However, as was already discussed in \cite{article1}, this method becomes less and less advantageous when the
value of $J$ increases ; only the $J=0$ and $J=1$ states have been computed so far, with an accuracy of 10$^{-14}$ a.u.
for all levels except for the last excited state and the $J=1$ states of H$_{2}^{+}$ and D$_{2}^{+}$ due to a higher
numerical instability. In this paper, we extend our method to the $J=2$ states and show that it remains advantageous with
respect to the usual methods relying on Hylleraas coordinates. For convenience, we have regrouped all our results in the
present paper ; all levels have been recomputed with the values of the most recent (2002) CODATA \cite{codata2002}, and we
also give their sensitivity with respect to the mass ratios. The accuracy for the $J=1$ states of H$_{2}^{+}$ and
D$_{2}^{+}$ has been improved to 10$^{-14}$ a.u. On the whole, 57 vibration-rotation levels of H$_{2}^{+}$, 79 levels of
D$_{2}^{+}$ and 65 levels of HD$^{+}$ have been computed with a precision of metrological interest (10$^{-12}$ a.u. or
better).

In the first section, we briefly recall the main features of our method (more details can be found in \cite{article1}).
The second section is devoted to the presentation of numerical results.

\section{Method of resolution}
\subsection{Hamiltonian}

Using centered Jacobi coordinates, the Hamiltonian of a three-body molecular ion with nuclear masses $M_{1}$ and $M_{2}$
can be written as
\begin{equation}
H = \frac{q^{2}}{4 \pi \epsilon_{0} a_{0}} \left( \frac{\mathbf{p}^{2}}{2}+\frac{1}{2 \mu_{12}
}\left(\mathbf{P}^{2}+\frac{\mathbf{p}^{2}}{4} \right)+\frac{\mathbf{p}.\mathbf{P}}{2 \mu_{0}}
-\frac{1}{\|\mathbf{R}/2-\mathbf{r}\|}- \frac{1}{\|\mathbf{R}/2+\mathbf{r}\|}+\frac{1}{R} \right), \label{hamiltonien}
\end{equation}
where $q$ is the electron charge, $a_{0}$ the atomic Bohr radius, $\mu_{12}=M_{1} M_{2}/m_{e}(M_{1}+M_{2})$ is the reduced
mass of the two nuclei in units of $m_{e}$, and $1/\mu_{0} = m_{e}(1/M_{1} - 1/M_{2})$. The dimensionless quantities
\textbf{R} and  \textbf{r} represent respectively the relative position of the two nuclei and the position of the electron
with respect to their center of mass. The quantities \textbf{P} and \textbf{p} are the conjugate momenta. The term in
\textbf{p}.\textbf{P} is the so-called symmetry breaking term, which vanishes for the homonuclear ions H$_{2}^{+}$ and
D$_{2}^{+}$.
\subsection{Structure of the wave functions}

The rotational invariance yields the following separation between angular and radial variables:
\begin{equation}
\Psi^{JM}(\mathbf{R},\mathbf{r}) = \sum_{T=-J}^{J} D^{J*}_{MT}(\psi, \theta, \phi) \; \Phi^{JM}_{T} (R,\rho,\zeta),
\end{equation}
where the radial coordinates $R$, $\rho$, $\zeta$ have been defined in \cite{article1}, $\psi$, $\theta$, $\phi$ are the
Euler angles, and $D^{J*}_{MT}$ are known angular functions related to the matrix elements of the rotation operator. As a
result, for a given value of $J$ and $M$, the wave functions are represented by $2J+1$ unknown radial functions.

Another symmetry of the Hamiltonian is the parity $\Pi$. Since the parity only affects the angular part of the wave
function, it is useful to introduce even and odd angular functions ; from the $2J+1$ functions $D^{J*}_{MT}$ one obtains
$J$ functions of parity $\Pi = (-1)^{J+1}$ and $J+1$ functions of parity $\Pi = (-1)^{J}$.

In this paper, we consider bound states that are supported by the first Born-Oppenheimer curve ($1s \sigma_{g}$) with a
total angular momentum $J \leq 2$. These states have the symmetries $S^{e}$, $P^{o}$, $D^{e}$ and the corresponding
separation between angular and radial variables is written, in the case of $M=0$ states:
\begin{eqnarray}
\mbox{S$^{e}$ states:} \hspace{0.5cm} \Psi^{00} (\mathbf{R},\mathbf{r})&=& \Phi^{0} (R,\rho,\zeta) \\
\mbox{P$^{o}$ states:} \hspace{0.5cm} \Psi^{10} (\mathbf{R},\mathbf{r})&=& D_{00}^{1*} (\psi, \theta, \phi) \;
\Phi^{0} (R,\rho,\zeta) \nonumber \\
&+& \frac{D_{0-1}^{1*}(\theta,\psi,\phi) - D_{01}^{1*}(\theta,\psi,\phi)} {\sqrt{2}} \; \Phi^{1} (R,\rho,\zeta) \\
&=& \cos \theta \; \Phi^{0} (R,\rho,\zeta) - \sin \theta \, \cos \phi \; \Phi^{1} (R,\rho,\zeta) \nonumber \\
\mbox{D$^{e}$ states:} \hspace{0.5cm} \Psi^{20} (\mathbf{R},\mathbf{r})&=& D_{00}^{2*} (\psi, \theta, \phi) \; \Phi^{0}
(R,\rho,\zeta) \nonumber \\
&+& \frac{D_{0-1}^{2*}(\theta,\psi,\phi) - D_{01}^{2*}(\theta,\psi,\phi)} {\sqrt{2}} \; \Phi^{1} (R,\rho,\zeta)
\nonumber\\
&+& \frac{D_{0-2}^{2*}(\theta,\psi,\phi) + D_{02}^{2*}(\theta,\psi,\phi)} {\sqrt{2}} \; \Phi^{2} (R,\rho,\zeta) \\
&=& \frac{3 \; \cos^{2} \theta - 1}{2} \; \Phi^{0} (R,\rho,\zeta) - \frac{\sqrt{3}}{2} \, \sin (2\theta) \, \cos \phi \;
\Phi^{1} (R,\rho,\zeta) \nonumber \\
&+& \frac{\sqrt{3}}{2} \, \sin^{2} \theta \, \cos(2 \phi) \; \Phi^{2} (R,\rho,\zeta) \nonumber
\end{eqnarray}
As a result, the radial part of the wave function is represented by $J+1$ radial functions $\Phi^{i} (R,\rho,\zeta)$.

\subsection{Factorization of the radial part}

The separation introduced in the previous section allows us to write down an effective Schr\"{o}dinger equation for the
radial part of the wave functions (for $J > 0$ it is in fact a system of coupled equations involving the $J+1$ radial
functions $\Phi^{i} (R,\rho,\zeta)$). In order to regularize the divergence introduced by the centrifugal terms (for $J
> 0$), it is necessary to factorize the radial wave functions. A general method of factorization was derived in
\cite{schwartz}. One obtains, for P$^{o}$ states :
\begin{eqnarray}
\Phi^{0} (R,\rho,\zeta) &=& \left( \zeta + \frac{R}{2} \right) \; F (R,\rho,\zeta) + \left( \zeta - \frac{R}{2} \right) \;
G(R,\rho,\zeta) \nonumber \\
\Phi^{1} (R,\rho,\zeta) &=& \rho \; F (R,\rho,\zeta) + \rho \; G(R,\rho,\zeta) \label{factorP}
\end{eqnarray}
and for D$^{e}$ states :
\begin{eqnarray}
\Phi^{0} (R,\rho,\zeta) &=& \left[ \left( \zeta + \frac{R}{2} \right)^{2} - \frac{\rho^{2}}{2} \right] \; F (R,\rho,\zeta)
+ \left[ \left( \zeta - \frac{R}{2} \right)^{2} - \frac{\rho^{2}}{2} \right] \; G(R,\rho,\zeta) \nonumber \\
&&+ \left[ \zeta^{2}-\frac{R^{2}}{4} - \frac{\rho^{2}}{2} \right] \; H(R,\rho,\zeta) \nonumber \\
\Phi^{1} (R,\rho,\zeta) &=& \sqrt{3} \, \rho \left( \zeta + \frac{R}{2} \right) \; F (R,\rho,\zeta) + \sqrt{3} \, \rho
\left(\zeta - \frac{R}{2} \right) \; G(R,\rho,\zeta) \nonumber \\
&&+ \sqrt{3} \, \rho \, \zeta \; H (R,\rho,\zeta) \nonumber \\
\Phi^{2} (R,\rho,\zeta) &=& \frac{\sqrt{3} \, \rho^{2}}{2} \; F (R,\rho,\zeta) + \frac{\sqrt{3} \, \rho^{2}}{2} \;
G(R,\rho,\zeta) + \frac{\sqrt{3} \, \rho^{2}}{2} \; H (R,\rho,\zeta) \label{factorD}
\end{eqnarray}
Using these expressions, we can write down a set of effective Schr\"{o}dinger equations for the radial functions $F,G,H$
appearing in (\ref{factorP}-\ref{factorD}), in which centrifugal terms of the type $1/R^{2}$ have disappeared. The
effective Hamiltonian $H_{eff}$ appearing in these equations can be found in \cite{article1} for P$^{o}$ states.

\subsection{Exchange symmetry}

In the cases of H$_{2}^{+}$ and D$_{2}^{+}$, we have an additional symmetry corresponding to the exchange of the two
nuclei. The wave functions are either symmetric or antisymmetric with respect to the exchange operator $P_{12}$. Like in
the atomic case, we will note here spatially symmetric (respectively antisymmetric) as singlets (respectively triplets).
The bound states of H$_{2}^{+}$ and D$_{2}^{+}$ that are considered in this paper have the following symmetries :
$^{1}$S$^{e}$, $^{3}$P$^{o}$ and $^{1}$D$^{e}$.

The effect of the exchange operator on radial functions is the transformation $\zeta \longrightarrow -\zeta$. Thus the
radial functions of singlet and triplet states differ by their behaviour, either symmetric or antisymmetric under the
transformation $\zeta \longrightarrow -\zeta$. For the states considered here, the radial functions have the following
properties :
\begin{eqnarray}
\mbox{$^{1}$S$^{e}$ states:} \hspace{0.5cm} F &=& \widetilde{F} \\
\mbox{$^{3}$P$^{o}$ states:} \hspace{0.5cm} G &=& -\widetilde{F} \\
\mbox{$^{1}$D$^{e}$ states:} \hspace{0.5cm} G &=& \widetilde{F} \; \mbox{and} \; H = \widetilde{H}
\end{eqnarray}
where $\widetilde{F} (R,\rho,\zeta) = F (R,\rho,-\zeta)$.

\subsection{Numerical implementation}

Even though the centrifugal terms have been eliminated, divergences in $1/r_{1}$, $1/r_{2}$ and $1/R$ remain due to the
Coulomb potential ($r_{1(2)}= \| \bf{R} \pm \bf{r}/2 \|$ are the distances between the electron and the two nuclei). These
divergences can be regularized through multiplication of the Schr\"{o}dinger equation by $r_{1} r_{2} R$. The
Schr\"{o}dinger equation is turned into a generalized eigenvalue problem, which is written as :
\begin{equation}
A \; |\Psi \rangle = E \; B \; |\Psi \rangle
\end{equation}
with $B = r_{1} r_{2} R$ and $A = B H_{eff}$.

Our numerical method to solve this problem has been explained in detail in \cite{article1}. It relies on the use of
perimetric radial coordinates, defined by:
\begin{eqnarray}
x&=&r_{1}+r_{2}-R  \nonumber \\
y&=&r_{1}-r_{2}+R   \\
z&=&-r_{1}+r_{2}+R,  \nonumber \label{coordonnees perimetriques}
\end{eqnarray}
and Sturmian basis functions in the $x,y,z$ coordinates
\begin{equation}
|n_{x}^{\alpha},n_{y}^{\beta},n_{z}^{\beta}\rangle = |n_{x}^{\alpha}\rangle \otimes |n_{y}^{\beta}\rangle \otimes
|n_{z}^{\beta}\rangle,
\end{equation}
where $|n_{u}^{\alpha}\rangle$ represents the function
\begin{equation}
\Phi_{n}(\alpha u)=\langle u|n^{\alpha}\rangle = (-1)^{n}\sqrt{\alpha}L_{n}^{\left(0\right)}\left(\alpha
u\right)e^{-\alpha u/2}.
\end{equation}
$n$ is a non-negative integer and $L_{n}^{(p)}$ the generalized Laguerre polynomials. $\alpha^{-1}$ is a length scale in
the $x$ direction, $\beta^{-1}$ in the $y$ and $z$ directions.

For H$_{2}^{+}$ and D$_{2}^{+}$, due to the exchange symmetry, some of the radial wave functions can be either symmetric
or antisymmetric with respect to the transformation $\zeta \longrightarrow -\zeta$, that is the exchange of $y$ and $z$.
In such cases we use a symmetrized or antisymmetrized basis :
\begin{equation}
|n_{x},n_{y},n_{z}\rangle^{\pm} = \frac{ |n_{x},n_{y},n_{z}\rangle \pm |n_{x},n_{z},n_{y}\rangle}{\sqrt{2}}
\end{equation}
To perform the numerical calculations, the basis is truncated at $n_{x} + n_{y} + n_{z} \leq N$ and $n_{x} \leq N_{x}$,
with $N_{x} \leq N$. If the basis is symmetrized (resp. antisymmetrized) we add the condition $n_{y} \leq n_{z}$ (resp.
$n_{y} < n_{z}$) which reduces the size of the basis by a factor of about 2. Since the radial part of the wave function is
represented by $J+1$ radial functions, the size of the matrices representing the $A$ and $B$ operators also depends on
$J$. The different cases are summarized in Table \ref{taillebase}.

\begin{table}
\center \small{
\begin{tabular}{|c|c|c|c|c|c|c|}
\hline
& \multicolumn{2}{|c|}{$J=0$} & \multicolumn{2}{|c|}{$J=1$} & \multicolumn{2}{|c|}{$J=2$}\\
\hline
& H$_{2}^{+}$, D$_{2}^{+}$ & HD$^{+}$ & H$_{2}^{+}$, D$_{2}^{+}$ & HD$^{+}$ & H$_{2}^{+}$, D$_{2}^{+}$ & HD$^{+}$ \\
\hline
radial wave function & $F = \widetilde{F}$ & $F$ & $F$ & $F$, $G$ & $F$, $H = \widetilde{H}$ & $F$, $G$, $H$ \\
\hline
basis size & $\sim N_{tot}/2$ & $N_{tot}$ & $N_{tot}$ & $2 \; N_{tot}$ & $\sim 3 N_{tot}/2$ & $3 N_{tot}$ \\
\hline
coupling rules & \multicolumn{2}{|c|}{$\left| \Delta n_{x,y,z} \right| \leq 2$,}& \multicolumn{2}{|c|}{$\left|
\Delta n_{x,y,z} \right| \leq 4$,} & \multicolumn{2}{|c|}{$\left| \Delta n_{x,y,z} \right| \leq 6$,} \\
& \multicolumn{2}{|c|}{\tiny {$\left| \Delta n_{x} \right| + \left| \Delta n_{y} \right| + \left| \Delta n_{z} \right|
\leq 3$}} & \multicolumn{2}{|c|}{\tiny {$\left| \Delta n_{x} \right| + \left| \Delta n_{y} \right| + \left| \Delta n_{z}
\right| \leq 5$}} & \multicolumn{2}{|c|}{\tiny{ $\left| \Delta n_{x} \right| + \left| \Delta n_{y} \right| + \left| \Delta
n_{z} \right| \leq 7$}} \\
\hline number of coupling rules & \multicolumn{2}{|c|}{57} & \multicolumn{2}{|c|}{450} & \multicolumn{2}{|c|}{1707} \\
\hline \multicolumn{7}{c}{} \\
\multicolumn{7}{c}{for \hspace{0.5cm} $N = 60$ \hspace{0.5cm} and \hspace{0.5cm} $N_{x} = 15$} \\
\hline basis size & 11964 & 23496 & 23496 & 46992 & 35460 & 70488 \\
\hline width & 834 & 1364 & 3031 & 4552 & 6422 & 8656 \\
\hline
\end{tabular}
} \caption{\label{taillebase} For all the computed levels, we have indicated in this table : the radial functions
representing the wave function, and (if necessary) their symmetry with respect to the exchange of $y$ and $z$ ; the basis
size as a function of $N_{tot}$, where $N_{tot}$ is the number of Sturmian functions $|n_{x},n_{y},n_{z}\rangle$ verifying
$n_{x} + n_{y} + n_{z} \leq N$ and $n_{x} \leq N_{x}$ ; the coupling rules between $|n_{x},n_{y},n_{z}\rangle$ and $|n_{x}
+ \Delta n_{x},n_{y} + \Delta n_{y},n_{z} + \Delta n_{z}\rangle$ and the number of coupling rules. Finally, for a typical
value of $N$ and $N_{x}$ we give the basis size and the width of the matrix.}
\end{table}

Because of their structure, all the terms in the Hamiltonian have strict coupling rules. $A$ and $B$ are then sparse band
matrices having exactly the same shape ; the order of the basis vectors is chosen in order to minimize their width around
the diagonal. The coupling rules and width are also reported in Table \ref{taillebase}; the fact that the number of
coupling rules increases with $J$ is due to the factorization, which involves polynomials of degree $J$ (see
Eqs.~(\ref{factorP})-(\ref{factorD})). As a result, with increasing $J$ the matrices become both larger and wider.

The analytical calculation of the matrix elements of the various contributions to the Hamiltonian has been performed using
the symbolic calculation language Mathematica 4. The results are directly output in double precision FORTRAN code. The
generalized eigenvalue problem is then diagonalized using the Lanczos algorithm. That gives the eigenvectors and
eigenvalues in the energy range of interest.

\section{Numerical results}

The energy levels of H$_{2}^{+}$, D$_{2}^{+}$ and HD$^{+}$ are given in Tables \ref{h2plus}, \ref{d2plus} and
\ref{hdplus}. The mass ratios are taken from the 2002 CODATA \cite{codata2002}: $M_{P}/m_{e}$ = 1836.15267261 and
$M_{D}/m_{e}$ = 3670.4829652. The atomic unit of energy is 219474.6313705 cm$^{-1}$. All the digits shown in these tables
are converged ; an accuracy of 10$^{-14}$ atomic unit, limited by the numerical noise, is achieved for most levels. Let us
stress that these results were obtained with quite reasonable computation resources, i.e. a single standard workstation
with 8 Go memory and double precision arithmetic.

When the accuracy is sufficient to be sensitive to the mass ratios, we have computed the normalized sensitivity $\lambda
\, \partial E/\partial \lambda$ of the energy levels to the variation of the relevant mass ratio(s), that is the
electron/nucleus mass ratio $\lambda = m_{e}/M_{P}$ for H$_{2}^{+}$ or HD$^{+}$ ($\lambda = m_{e}/M_{D}$ for D$_{2}^{+}$)
and the ratio of the nuclear masses $\mu  = M_{P}/M_{D}$ for HD$^{+}$, following the method described in \cite{schiller}.
The sensitivities are given in Tables \ref{pentesh2plus}, \ref{pentesd2plus} and \ref{penteshdplus}. Again, all the digits
shown in these tables are converged ; they are in full agreement with the values published in \cite{schiller}. From these
tables it is possible to calculate the energy levels for any value of the mass ratios within their present area of
uncertainty (in the 2002 CODATA the relative uncertainty is 4.6 10$^{-10}$ for $m_{e}/M_{P}$, 4.8 10$^{-10}$ for
$m_{e}/M_{D}$ and 2 10$^{-10}$ for $M_{P}/M_{D}$) while maintaining the same accuracy.

The precision for the $J=1$ states of H$_{2}^{+}$ and D$_{2}^{+}$ has been improved from 10$^{-11}$ to 10$^{-14}$ a.u.
with respect to earlier results \cite{article1}. As discussed in \cite{article1}, the increased numerical noise was due to
round off errors that accumulate more rapidly during the Lanczos steps, because the matrices are ill-conditioned. However,
for a given eigenvalue $E_{n}$ the accuracy of the result can be improved by solving the problem $(A - \lambda B)\; |\Psi
\rangle = E \; B \; |\Psi \rangle$ with $\lambda$ close to $E_{n}$. In this way, the eigenvalue is converged during the
first few Lanczos steps and the round off errors are greatly reduced. Of course, the downside is that a new matrix has to
be diagonalized for each level, which multiplies the computation time by the number of computed levels.

For $J=2$ levels, the conditioning properties of the matrices are even worse than for $J=1$ levels. It is not surprising
that the conditioning should worsen when $J$ increases, because their matrix elements are polynomials in $n_{x}$, $n_{y}$,
$n_{z}$, the degree of which increases with $J$ since the degree of the polynomials appearing in the factorization
increases (see expressions (\ref{factorP}), (\ref{factorD})). However, for most levels it is still possible to achieve an
accuracy of 10$^{-14}$ a.u. Only for the last few levels (for example $v=16$ and $v=17$ in H$_{2}^{+}$, or $v=20$ to $24$
in D$_{2}^{+}$) does the numerical accuracy drop to 10$^{-13}$, because of the increased basis size. On the whole, for
most levels the precision has been improved by 4 or 5 orders of magnitude with respect to earlier works
\cite{mosshdplus,mossh2plus,mossd2plus} except for the first vibrational levels of HD$^{+}$ ($v=0$ to $v=4$) which had
been computed with 10$^{-14}$ or 10$^{-15}$ accuracy in \cite{schiller}. We have checked that our results are in full
agreement with those of that reference.

Finally, let us note that for $J=0$ and $J=1$, only the last vibrational level (or the last two levels) could not be
converged to 10$^{-14}$ accuracy due to the limit in memory size. $J=2$ levels demand more memory (since the matrices are
both larger and wider, see Table \ref{taillebase}) which is why the level of convergence drops for the last three or four
vibrational levels. Whenever the accuracy of our calculations did not reach 10$^{-9}$-10$^{-10}$ a.u. we have reported the
more precise results obtained by R. E. Moss \cite{mosshdplus,mossh2plus,mossd2plus}; although these results have been
obtained with the 1986 CODATA values of the mass ratios, all the digits remain valid because the sensitivity of the last
vibrational levels is very low.

All these elements make it clear that our method becomes less and less convenient when $J$ increases. It may be possible
to extend the calculations to $J=3$ states, but then two problems will occur: \\
- the limit in memory will prevent the calculation of a larger number of levels ; \\
- conditioning problems will become more severe ; unless an efficient method of pre-conditioning of the matrices can be
devised, quadruple precision arithmetic is likely to be compulsory in this case which imposes new constraints in terms of
memory and computation time.

\section{Conclusion}

We have shown that the use of perimetric coordinates allows to obtain high accuracy results for the nonrelativistic
rotation-vibration energies of the hydrogen molecular ion and its isotopes, up to $J=2$ ; our results suggest that this is
probably the last value of $J$ for which this method is advantageous with respect to usual methods relying on Hylleraas
coordinates. The very accurate wavefunctions that we obtain enable the calculation of relativistic and radiative
corrections with an improved precision, which is in progress. These results extend the range of optical transitions on
which high-precision measurements can be compared to theoretical predictions for metrological applications.

\ack

The authors wish to thank D Delande and B Gr\'emaud for fruitful discussions and also for providing us the Lanczos
diagonalization code. Laboratoire Kastler Brossel de l'Universit\'e Pierre et Marie Curie et de l'Ecole Normale
Sup\'erieure is UMR 8552 du CNRS.


\newpage

\begin{table}
\center
\begin{tabular}{|c|l|l|l|}
\hline
v/J & \multicolumn{1}{|c|}{0} & \multicolumn{1}{|c|}{1} & \multicolumn{1}{|c|}{2} \\
\hline
  0 & -0.597 139 063 079 39 & -0.596 873 738 784 71 & -0.596 345 205 489 39 \\
  1 & -0.587 155 679 096 19 & -0.586 904 320 919 19 & -0.586 403 631 528 69 \\
  2 & -0.577 751 904 415 08 & -0.577 514 034 057 45 & -0.577 040 237 163 02 \\
  3 & -0.568 908 498 730 86 & -0.568 683 708 260 19 & -0.568 235 992 971 58 \\
  4 & -0.560 609 220 849 67 & -0.560 397 171 400 29 & -0.559 974 864 820 05 \\
  5 & -0.552 840 749 896 55 & -0.552 641 171 550 16 & -0.552 243 738 618 17 \\
  6 & -0.545 592 650 993 83 & -0.545 405 343 957 06 & -0.545 032 389 906 00 \\
  7 & -0.538 857 386 967 41 & -0.538 682 224 240 87 & -0.538 333 500 061 38 \\
  8 & -0.532 630 379 356 27 & -0.532 467 311 197 56 & -0.532 142 722 733 45 \\
  9 & -0.526 910 124 016 32 & -0.526 759 184 659 29 & -0.526 458 806 281 70 \\
 10 & -0.521 698 369 014 24 & -0.521 559 686 353 44 & -0.521 283 780 673 62 \\
 11 & -0.517 000 365 278 75 & -0.516 874 174 628 27 & -0.516 623 220 542 74 \\
 12 & -0.512 825 203 145 56 & -0.512 711 866 864 11 & -0.512 486 599 658 18 \\
 13 & -0.509 186 248 368 29 & -0.509 086 284 367 52 & -0.508 887 754 205 98 \\
 14 & -0.506 101 680 968 76 & -0.506 015 805 446 04 & -0.505 845 465 600 80 \\
 15 & -0.503 595 084 999 22 & -0.503 524 279 325 77 & -0.503 384 125 722 45 \\
 16 & -0.501 695 773 387 03 & -0.501 641 393 537 72 & -0.501 534 197 631 3  \\
 17 & -0.500 437 040 460 15 & -0.500 400 984 505 45 & -0.500 330 669 647 8  \\
 18 & -0.499 837 432 030 23 & -0.499 821 792 336 03 & -0.499 792 794 5      \\
 19 & -0.499 731 230 649 2  & -0.499 728 846 9$^{a}$&\multicolumn{1}{|c|}{*}\\
\hline
\end{tabular}
\caption{\label{h2plus} Energies of the $^{1}$S$^{e}$, $^{3}$P$^{o}$ and $^{1}$D$^{e}$ bound levels of the H$_{2}^{+}$
molecular ion below the first dissociation limit, in atomic units. The star indicates that there is no bound level. $^{a}$
result taken from Ref. \cite{mossh2plus}. The first dissociation limit of H$_{2}^{+}$ is $-1/2 \; (1+m_{e}/M_{P})^{-1}$ =
-0.499 727 839 712 26 a.u.}
\end{table}

\begin{table}
\center
\begin{tabular}{|c|c|c|c|}
\hline
v/J & 0 & 1 & 2 \\
\hline
  0 & 0.284657 & 0.310779 & 0.362604 \\
  1 & 0.753846 & 0.777901 & 0.825615 \\
  2 &  1.16667 &  1.18874 &  1.23253 \\
  3 &  1.52570 &  1.54587 &  1.58586 \\
  4 &  1.83293 &  1.85126 &  1.88757 \\
  5 &  2.08979 &  2.10631 &  2.13902 \\
  6 &  2.29711 &  2.31184 &  2.34099 \\
  7 &  2.45516 &  2.46810 &  2.49368 \\
  8 &  2.56356 &  2.57469 &  2.59667 \\
  9 &  2.62128 &  2.63056 &  2.64884 \\
 10 &  2.62654 &  2.63389 &  2.64833 \\
 11 &  2.57673 &  2.58205 &  2.59241 \\
 12 &  2.46825 &  2.47138 &  2.47738 \\
 13 &  2.29639 &  2.29712 &  2.29831 \\
 14 &  2.05510 &  2.05313 &  2.04890 \\
 15 &  1.73686 &  1.73178 &  1.72127 \\
 16 &  1.33316 &  1.32434 &  1.30625 \\
 17 & 0.838586 & 0.825125 & 0.797542 \\
 18 & 0.293354 & 0.276164 & -        \\
 19 & 0.043008 & -        & *        \\
\hline
\end{tabular}
\caption{\label{pentesh2plus} Derivatives $10^{2} \, \lambda \, \partial E/\partial \lambda$ of the energies of
H$_{2}^{+}$ with respect to the electron/proton mass ratio $\lambda$ (in atomic units). The hyphens correspond to levels
for which the accuracy of the calculation is not sufficient to be sensitive to a change of the recommended value of the
proton/electron mass ratio.}
\end{table}

\begin{table}
\center
\begin{tabular}{|c|l|l|l|}
\hline
v/J & \multicolumn{1}{|c|}{0} & \multicolumn{1}{|c|}{1} & \multicolumn{1}{|c|}{2} \\
\hline
  0 & -0.598 788 784 304 46 & -0.598 654 873 192 49 & -0.598 387 585 778 48 \\
  1 & -0.591 603 121 831 23 & -0.591 474 211 454 95 & -0.591 216 909 547 45 \\
  2 & -0.584 712 206 896 08 & -0.584 588 169 503 36 & -0.584 340 598 262 38 \\
  3 & -0.578 108 591 284 75 & -0.577 989 311 807 81 & -0.577 751 241 739 67 \\
  4 & -0.571 785 598 461 03 & -0.571 670 974 249 74 & -0.571 442 200 677 76 \\
  5 & -0.565 737 302 734 64 & -0.565 627 243 389 04 & -0.565 407 586 123 15 \\
  6 & -0.559 958 513 978 72 & -0.559 852 941 284 34 & -0.559 642 244 496 68 \\
  7 & -0.554 444 767 877 07 & -0.554 343 615 849 03 & -0.554 141 748 167 64 \\
  8 & -0.549 192 321 773 66 & -0.549 095 536 818 38 & -0.548 902 391 656 91 \\
  9 & -0.544 198 156 295 12 & -0.544 105 697 502 03 & -0.543 921 193 650 73 \\
 10 & -0.539 459 983 025 35 & -0.539 371 822 605 81 & -0.539 195 905 118 31 \\
 11 & -0.534 976 258 632 93 & -0.534 892 382 529 17 & -0.534 725 023 951 62 \\
 12 & -0.530 746 205 989 53 & -0.530 666 614 684 77 & -0.530 507 816 690 08 \\
 13 & -0.526 769 842 975 10 & -0.526 694 552 546 50 & -0.526 544 348 058 54 \\
 14 & -0.523 048 019 841 42 & -0.522 977 063 312 93 & -0.522 835 519 236 33 \\
 15 & -0.519 582 466 193 55 & -0.519 515 895 267 28 & -0.519 383 115 983 02 \\
 16 & -0.516 375 848 821 83 & -0.516 313 736 098 48 & -0.516 189 867 950 32 \\
 17 & -0.513 431 841 702 04 & -0.513 374 283 548 56 & -0.513 259 520 643 92 \\
 18 & -0.510 755 209 287 13 & -0.510 702 329 583 95 & -0.510 596 921 385 54 \\
 19 & -0.508 351 903 244 84 & -0.508 303 858 362 06 & -0.508 208 119 790 46 \\
 20 & -0.506 229 169 712 90 & -0.506 186 155 255 51 & -0.506 100 480 422 93 \\
 21 & -0.504 395 655 120 05 & -0.504 357 915 292 30 & -0.504 282 796 650 0  \\
 22 & -0.502 861 471 598 77 & -0.502 829 312 510 71 & -0.502 765 368 237 6  \\
 23 & -0.501 638 094 606 39 & -0.501 611 903 260 41 & -0.501 559 916 810 6  \\
 24 & -0.500 737 626 271 87 & -0.500 717 894 503 72 & -0.500 678 865 993 5  \\
 25 & -0.500 169 272 837 37 & -0.500 156 588 424 55 & -0.500 131 723 49     \\
 26 & -0.499 919 155 009 21 & -0.499 913 606 74     & -0.499 903 071 2$^{a}$\\
 27 & -0.499 868 405 486    & -0.499 866 917 5$^{a}$& -0.499 864 451 9$^{a}$\\
\hline
\end{tabular}
\caption{\label{d2plus} Same as Table \ref{h2plus}, for the D$_{2}^{+}$ molecular ion. $^{a}$ result taken from Ref.
\cite{mossd2plus}. The first dissociation limit of D$_{2}^{+}$ is -0.499 863 815 247 21 a.u.}
\end{table}

\begin{table}
\center
\begin{tabular}{|c|c|c|c|}
\hline
v/J & 0 & 1 & 2 \\
\hline
  0 & 0.197274 & 0.210525  & 0.236919 \\
  1 & 0.541450 & 0.553958  & 0.578871 \\
  2 & 0.856748 & 0.868537  & 0.892017 \\
  3 & 1.14421  &  1.15530  &  1.17739 \\
  4 & 1.40471  &  1.41513  &  1.43586 \\
  5 & 1.63901  &  1.64876  &  1.66817 \\
  6 & 1.84769  &  1.85679  &  1.87490 \\
  7 & 2.03120  &  2.03966  &  2.05649 \\
  8 & 2.18987  &  2.19769  &  2.21325 \\
  9 & 2.32384  &  2.33103  &  2.34533 \\
 10 & 2.43313  &  2.43968  &  2.45272 \\
 11 & 2.51759  &  2.52351  &  2.53527 \\
 12 & 2.57682  &  2.58219  &  2.59266 \\
 13 & 2.61062  &  2.61523  &  2.62438 \\
 14 & 2.61801  &  2.62194  &  2.62973 \\
 15 & 2.59819  &  2.60142  &  2.60780 \\
 16 & 2.55004  &  2.55253  &  2.55744 \\
 17 & 2.47217  &  2.47388  &  2.47724 \\
 18 & 2.36289  &  2.36378  &  2.36549 \\
 19 & 2.22021  &  2.22021  &  2.22013 \\
 20 & 2.04174  &  2.04077  &  2.03874 \\
 21 & 1.82476  &  1.82270  &  1.81852 \\
 22 & 1.56620  &  1.56293  &  1.55630 \\
 23 & 1.26306  &  1.25840  &  1.24896 \\
 24 & 0.913898 & 0.907605  & 0.894869 \\
 25 & 0.526414 & 0.518342  & -        \\
 26 & 0.168515 & -         & -        \\
 27 & 0.033736 & -         & -        \\
\hline
\end{tabular}
\caption{\label{pentesd2plus} Derivatives $10^{2} \, \lambda \, \partial E/\partial \lambda$ of the energies of
D$_{2}^{+}$ with respect to the electron/deuteron mass ratio $\lambda$ (in atomic units). The hyphens correspond to levels
for which the accuracy of the calculation is not sufficient to be sensitive to a change of the recommended value of the
deuteron/electron mass ratio.}
\end{table}

\begin{table}
\center
\begin{tabular}{|c|l|l|l|}
\hline
v/J & \multicolumn{1}{|c|}{0} & \multicolumn{1}{|c|}{1} & \multicolumn{1}{|c|}{2} \\
\hline
  0 & -0.597 897 968 609 03 & -0.597 698 128 192 21 & -0.597 299 643 351 78 \\
  1 & -0.589 181 829 556 96 & -0.588 991 111 992 04 & -0.588 610 829 389 79 \\
  2 & -0.580 903 700 218 37 & -0.580 721 828 120 93 & -0.580 359 195 199 88 \\
  3 & -0.573 050 546 551 87 & -0.572 877 277 094 21 & -0.572 531 810 325 97 \\
  4 & -0.565 611 042 076 81 & -0.565 446 166 277 57 & -0.565 117 449 763 74 \\
  5 & -0.558 575 520 825 56 & -0.558 418 863 258 87 & -0.558 106 548 173 82 \\
  6 & -0.551 935 948 956 82 & -0.551 787 367 908 52 & -0.551 491 172 840 00 \\
  7 & -0.545 685 915 292 93 & -0.545 545 303 409 43 & -0.545 265 015 689 56 \\
  8 & -0.539 820 641 545 82 & -0.539 687 927 044 13 & -0.539 423 405 211 03 \\
  9 & -0.534 337 013 561 27 & -0.534 212 162 100 65 & -0.533 963 339 694 69 \\
 10 & -0.529 233 635 566 11 & -0.529 116 652 925 66 & -0.528 883 543 915 03 \\
 11 & -0.524 510 910 171 74 & -0.524 401 845 942 75 & -0.524 184 552 204 94 \\
 12 & -0.520 171 147 776 22 & -0.520 070 100 375 34 & -0.519 868 821 861 27 \\
 13 & -0.516 218 709 961 71 & -0.516 125 833 423 20 & -0.515 940 881 945 48 \\
 14 & -0.512 660 192 252 70 & -0.512 575 705 505 94 & -0.512 407 523 608 73 \\
 15 & -0.509 504 651 335 11 & -0.509 428 851 096 13 & -0.509 278 038 350 31 \\
 16 & -0.506 763 877 817 30 & -0.506 697 156 979 60 & -0.506 564 507 652 98 \\
 17 & -0.504 452 698 864 25 & -0.504 395 573 710 86 & -0.504 282 132 868 4  \\
 18 & -0.502 589 233 789 25 & -0.502 542 386 290 66 & -0.502 449 537 962 0  \\
 19 & -0.501 194 799 118 31 & -0.501 159 147 958 76 & -0.501 088 766 606    \\
 20 & -0.500 292 454 206 62 & -0.500 269 323 125 65 & -0.500 224 158 4      \\
 21 & -0.499 910 361 470 03 & -0.499 902 783 135 5  & -0.499 889 109 $^{a}$ \\
 22 & -0.499 865 778 5      & -0.499 864 342 $^{a}$ &\multicolumn{1}{|c|}{*}\\
\hline
\end{tabular}
\caption{\label{hdplus} Energies of the S$^{e}$, P$^{o}$ and D$^{e}$ bound levels of the HD$^{+}$ molecular ion. The star
indicates that there is no bound level. $^{a}$result taken from Ref.\cite{mosshdplus}. The two dissociation limits of
HD$^{+}$ are HD$^{+} \longrightarrow$D+H$^{+}$ (given in the caption of Table \ref{d2plus}) and HD$^{+}
\longrightarrow$H+D$^{+}$ (given in the caption of Table \ref{h2plus}).}
\end{table}

\begin{table}
\center
\begin{tabular}{|c|c|c|c|c|c|c|}
\hline
v/J & \multicolumn{2}{|c|}{0} & \multicolumn{2}{|c|}{1} & \multicolumn{2}{|c|}{2} \\
\hline
    & $10^{2} \, \lambda \, \partial E / \partial \lambda$
    & $10^{2} \, \mu \, \partial E / \partial \mu$
    & $10^{2} \, \lambda \, \partial E / \partial \lambda$
    & $10^{2} \, \mu \, \partial E / \partial \mu$
    & $10^{2} \, \lambda \, \partial E / \partial \lambda$
    & $10^{2} \, \mu \, \partial E / \partial \mu$ \\
\hline
  0 & 0.244262 & 0.081468 & 0.263983 & 0.088045 & 0.303187 & 0.101120 \\
  1 & 0.657496 & 0.219284 & 0.675865 & 0.225410 & 0.712375 & 0.237586 \\
  2 &  1.02799 & 0.342846 &  1.04506 & 0.348538 &  1.07898 & 0.359851 \\
  3 &  1.35753 & 0.452749 &  1.37334 & 0.458023 &  1.40476 & 0.468501 \\
  4 &  1.64756 & 0.549475 &  1.66215 & 0.554342 &  1.69114 & 0.564011 \\
  5 &  1.89920 & 0.633398 &  1.91260 & 0.637868 &  1.93922 & 0.646744 \\
  6 &  2.11325 & 0.704785 &  2.12548 & 0.708864 &  2.14975 & 0.716960 \\
  7 &  2.29018 & 0.763794 &  2.30125 & 0.767485 &  2.32321 & 0.774808 \\
  8 &  2.43014 & 0.810472 &  2.44004 & 0.813775 &  2.45968 & 0.820325 \\
  9 &  2.53291 & 0.844748 &  2.54165 & 0.847661 &  2.55895 & 0.853431 \\
 10 &  2.59793 & 0.866431 &  2.60546 & 0.868946 &  2.62039 & 0.873922 \\
 11 &  2.62419 & 0.875192 &  2.63050 & 0.877298 &  2.64297 & 0.881457 \\
 12 &  2.61027 & 0.870553 &  2.61531 & 0.872233 &  2.62523 & 0.875541 \\
 13 &  2.55423 & 0.851869 &  2.55792 & 0.853100 &  2.56516 & 0.855512 \\
 14 &  2.45357 & 0.818302 &  2.45582 & 0.819055 &  2.46018 & 0.820508 \\
 15 &  2.30510 & 0.768797 &  2.30580 & 0.769029 &  2.30703 & 0.769442 \\
 16 &  2.10490 & 0.702043 &  2.10387 & 0.701702 &  2.10166 & 0.700964 \\
 17 &  1.84818 & 0.616454 &  1.84522 & 0.615466 &  1.83911 & 0.613429 \\
 18 &  1.52938 & 0.510190 &  1.52417 & 0.508453 &  1.51352 & 0.504907 \\
 19 &  1.14272 & 0.381394 &  1.13480 & 0.378759 &  1.11925 & 0.372827 \\
 20 & 0.684174 & 0.229190 & 0.672638 & 0.225385 & -        & -        \\
 21 & 0.142706 & 0.054888 & 0.128607 & 0.050433 & -        & -        \\
 22 & -        & -        & -        & -        & *        & *        \\
\hline
\end{tabular}
\caption{\label{penteshdplus} Derivatives of the energies of HD$^{+}$ with respect to the electron/proton and
proton/deuteron mass ratios, respectively noted $\lambda$ and $\mu$(in atomic units). The hyphens correspond to levels for
which the accuracy of the calculation is not sufficient to be sensitive to a change of the recommended value of the mass
ratios.}
\end{table}

\end{document}